\documentclass[pre,showpacs, preprintnumbers ,twocolumn]{revtex4}

\usepackage{graphicx}

\usepackage{amsmath}
\usepackage{amssymb}
\usepackage{mathrsfs}
\usepackage{bm}

\def\be{\begin{equation}}
\def\ee{\end{equation}}
\def\beu{\begin{equation*}}
\def\eeu{\end{equation*}}

\def\bse{\begin{subequations}}
\def\ese{\end{subequations}}

\def\Ez{\varepsilon}
\def\wp{\omega_p}

\def\En{\mathcal{E}}
\def\Ham{\mathcal{H}}
\def\action{J}
\def\act{\mathcal{I}}

\providecommand{\unitvec}[1]{\hat{\boldsymbol{#1}}}

\DeclareMathOperator{\ellipticE}{\mathscr{E}}
\DeclareMathOperator{\ellipticK}{\mathscr{K}}

\begin{document}

\title{Self-consistent Langmuir waves in resonantly driven thermal plasmas}
\author{R.R. Lindberg}
\email{rlindberg@berkeley.edu}
\author{A.E. Charman}
\author{J.S. Wurtele}
\affiliation{Department of Physics, University of California, Berkeley, Berkeley, CA 94720, USA\\
Center for Beam Physics, Lawrence Berkeley National Laboratory, USA}
\date{\today}

\begin{abstract}
The longitudinal dynamics of a resonantly driven Langmuir wave are analyzed in the limit that the growth of the electrostatic wave is slow compared to the bounce frequency.  Using simple physical arguments, the nonlinear distribution function is shown to be nearly gaussian in the canonical particle action, with a slowly evolving mean and fixed variance.  Self-consistency with the electrostatic potential provide the basic properties of the nonlinear distribution function including a frequency shift that agrees well with driven, electrostatic particle simulations.  This extends earlier work on nonlinear Langmuir waves by Morales and O'Neil [G.J.~Morales and T.M.~O'Neil, Phys.~Rev.~Lett.~\textbf{28}, 417 (1972)], and could form the basis of a reduced kinetic treatment of Raman backscatter in a plasma.
\end{abstract}
\maketitle

\section{Introduction}
Much of the progress in understanding Langmuir waves has been from the linear viewpoint, obtained by assuming that the perturbation of the plasma from its (Maxwellian) equilibrium is ``sufficiently small'' such that second order terms in the perturbation may be neglected.  Under these conditions, one can derive the normal modes of the distribution function (the singular Case-Van Kampen modes \cite{VanKampen:1955, Case:1959}) or, alternatively, the Landau damped ``modes'' of the electric field \cite{Landau:1946, Jackson:1959}.  A basic result of these linear analyses is that smooth, electrostatic perturbations tend to zero through the process of Landau damping.

That such damping is not universal was first pointed out by Bernstein, Green, and Kruskal (BGK) 
\cite{BernsteinGreeneKruskal:1957}, who included the particles trapped in the electrostatic wave to formulate nonlinear distribution functions that give rise to time-independent electrostatic disturbances.  Explicit constructions of sinusoidal, small-amplitude BGK waves were later derived by Holloway and Dorning \cite{HollowayDorning:1991}, in which they showed that even arbitrarily small amplitude waves can exist  without being Landau damped.  BGK distributions are generally functions of the conserved particle energy $H=\tfrac{1}{2}m_e v^2 - e\Phi(z)$ whose charge perturbation self-consistently generates the electrostatic potential $\Phi(z)$ via Poisson's equation.  Thus, BGK distributions are static solutions to the 1D Vlasov-Poisson system
\bse		\label{eqn:VlasovPoisson}
\begin{align}
  \frac{d}{dt}f(v,z;t) &= \frac{\partial f}{\partial t} + v \frac{\partial f}{\partial z} + 
  	e\frac{\partial\Phi}{\partial z} \frac{\partial f}{\partial v} = 0	\label{eqn:Vlasov} \\
  \frac{\partial^2}{\partial z^2}\Phi(z) &= 4\pi e\!\int \! dv\; f(v,z;t) - 4\pi e n_i(z),   \label{eqn:Poisson}
\end{align}
\ese
where $f$ is the electron distribution function and we consider the ions to have a time-independent density $n_i(z)$.

In this paper, we introduce and characterize nonlinear Langmuir wave solutions to the Vlasov-Poisson system (\ref{eqn:VlasovPoisson}) that are naturally occurring BGK-like waves.  These waves (and the distribution functions that generate them) have particular relevance to laser-plasma physics, in that they dynamically arise as kinetic, nonlinear Langmuir waves in systems that are weakly driven on or near resonance.  To obtain these solutions, we use the canonical action-angle coordinates, finding that the plasma is well-described by a simplified distribution function that is gaussian in the canonical action.  In this way, we obtain near-equilibrium solutions that approximate the fully time-dependent distribution function when the resonant forcing is small.  While these notions may be reminiscent of adiabatic theory, we do not invoke adiabatic invariance, since we imagine that the plasma is resonantly driven.  Our calculation is more in the spirit of an averaged theory, in that the dynamical dependence on the canonical angle is ignored on the grounds of rapid phase-mixing in the Langmuir wave, while the particle action evolves self-consistently.

Because these nonlinear, kinetic Langmuir waves arise naturally in slowly driven systems, their bulk properties can be used to illuminate basic plasma processes and obtain reduced descriptions of complex phenomena.  For example, the nonlinear frequency shift of the thermal Langmuir resonance is an important quantity in any reduced model of Raman scattering in plasma \cite{Berger_et_al:1998, RoseRussell:2001, Vu_et_al:2002, Divol_et_al:2003}, and our results extend those of Morales and O'Neil \cite{MoralesONeil:1972} to colder plasmas and larger electrostatic potentials $\Phi(z)$.  We leave such an implementation in a Langmuir envelope code to future work.

In Sec.~\ref{section:pendulum} we first present the single particle equations of motion and then show that for a weakly driven system, the particles move in an essentially sinusoidal potential.  We proceed by reviewing the relevant pendulum dynamics and action-angle coordinates.  In 
Sec.~\ref{section:distribution} we introduce the plasma distribution function, and use a few simple, physically motivated assumptions to show that a slowly excited plasma is well-represented by a distribution function that is gaussian in the canonical action with a fixed variance.  Next, in Sec.~\ref{section:BGK}, we use Coulomb's law and the demands of self-consistency to derive the functional relationship between the mean action and the amplitude of the potential.  This fully specifies the distribution function, from which we then extract the natural frequency of the BGK-type wave.  We compare the mean action and frequency of these nonlinear Langmuir waves to those obtained from self-consistent particle simulations in Sec.~\ref{section:simulation} for thermal plasmas with 
$0.4 \le k\lambda_D \le 0.2$, where $\lambda_D \equiv v_{\text{th}}/\omega_p$.  Some concluding remarks and possible applications are given in Sec.~\ref{section:conclusions}.

\section{Single particle equations of motion: the driven pendulum}	\label{section:pendulum}

In this section we present the single particle equations and derive the canonical action-angle variables relevant to a weakly-driven plasma wave.  These familiar results are subsequently used in 
Sec.~\ref{section:distribution} to motivate our simplified, fluid-like characterization of the electron distribution function for a slowly growing Langmuir wave.

In what follows we ignore transverse variation, assuming that the dominant dynamics are along the longitudinal axis $z$.  We ignore the motion of the background ions, and furthermore assume that the longitudinal force on the electrons can be divided into two components: the first is given by an external driving potential $V(z,t) \sin(\omega t + k z)$, which could be, for example, a ponderomotive force; the second arises from the self-consistent electrostatic potential $\Phi(z,t)$ of the plasma electrons.  Thus, Newton's equation of motion for the longitudinal electron coordinate $z(t)$ is given by
\begin{align}
  \frac{d^2}{dt^2} z(t) &= \frac{\partial}{\partial z}\Big[ e\Phi(z,t) -
  	V(z,t) \sin(\omega t + k z) \Big].		\label{eqn:Newton1}
\end{align}
Note that for simplicity we assume the dynamics to be nonrelativistic, requiring that the potentials remain sufficiently small such that $\left\lvert e\Phi \right\rvert,\, \left\lvert V\right\rvert \ll m_e c^2\,$; for a nonrelativistic external drive $V$, the electrostatic potential will satisfy this relation for all time if the nominal phase velocity is much less than the speed of light, $(\omega/k)^2 \ll c^2$.

To simplify (\ref{eqn:Newton1}), we assume that the amplitude of the external potential $V(z,t)$ has a slow spatiotemporal variation with respect to its carrier phase, so that
\begin{align}
  \frac{\partial}{\partial z}\ln \left\lvert V(z,t) \right\rvert &\ll k &
    	\frac{\partial}{\partial t} \ln \left\lvert V(z,t) \right\rvert \ll \omega.		\label{eqn:Eik}
\end{align}
The nearly-periodic external drive sets a natural spatial length of the driven self-consistent electrostatic potential, for which $\Phi(z,t)$ can be Fourier expanded in terms of dimensionless Eikonal amplitudes:
\be
  \Phi(z,t) = \frac{m_e \wp^2}{e k^2}\sum_{n=1}^\infty 
  	\phi_n(z,t) \cos[n(\omega t + k z) + \xi_n(z,t)].		\label{eqn:FourPhi}
\ee
For later convenience, we use a cosine series with the Langmuir phase shifts given by $\xi_n(z,t)$, and the normalization is chosen such that, for all $z$ and $t$,
\beu
  \left\lvert \phi_n(z,t) \right\rvert \le \frac{1}{n^2}.
\eeu
Furthermore, since $\Phi(z, t)$ is excited by the slowly varying potential $V(z,t)$, the Eikonal conditions (\ref{eqn:Eik}) imply that the amplitudes $\phi_n(z,t)$ and phase shifts $\xi_n(z,t)$ are similarly slowly varying.  We assume that the plasma is not highly nonlinear, so that the potential is adequately described by its first $(n=1)$ harmonic.  We have found that this assumption is not overly restrictive: for 
$\phi_1$ as large as one-half simulations indicate that $\phi_2 \sim \tfrac{1}{10}\phi_1$.  Finally, we assume that the space-charge force from the plasma bulk is dominant, i.e.,  $\left\lvert V(z,t) \right\rvert \ll \left\lvert \phi_1(z,t)\right\rvert$.   With these assumptions, Newton's equation (\ref{eqn:Newton1}) simplifies to
\be
  \frac{d^2}{dt^2} z_j(t) \approx -\frac{\wp^2}{k}\phi_1(z, t) \sin[\omega t + k z_j + \xi_1(z,t)].
  	\label{eqn:Newton2}
\ee

To express (\ref{eqn:Newton2}) as a Hamiltonian system appropriate for action-angle variables, we introduce the dimensionless time $\tau \equiv \wp t$, the scaled (linear) frequency 
$\omega_L \equiv \omega/\wp$, and the dimensionless coordinates given by the phase in the electrostatic wave $\theta$ and its corresponding canonical momentum $p$:
\begin{align}
  \theta &\equiv \omega t + k z_j + \xi_1 & p & \equiv \dot{\theta} - \dot{\xi}_1 = k \dot{z}_j + 
  	\omega_L,	\label{eqn:pqDefs}
\end{align}
with the over-dot understood to denote the normalized time derivative $\tfrac{d}{d\tau} \equiv \tfrac{1}{\wp}\tfrac{d}{dt}$.  Defining the frequency shift $\delta\omega \equiv \dot{\xi_1}$, 
(\ref{eqn:Newton2}) becomes
\begin{align}
  \dot{\theta} &= p + \delta\omega(\tau)	& \dot{p} &= -\phi_1(\tau)\sin\theta.	\label{eqn:EOMqp}
\end{align}
The system (\ref{eqn:EOMqp}) can be obtained from the pendulum Hamiltonian
\be
  \Ham(p, \theta; \tau) = \frac{1}{2}\left[p + \delta\omega(\tau)\right]^2 + 2\phi_1(\tau)\sin^2(\theta/2).
  	\label{eqn:PendHam}
\ee

\begin{figure}
\begin{centering}
\includegraphics[scale=1.2]{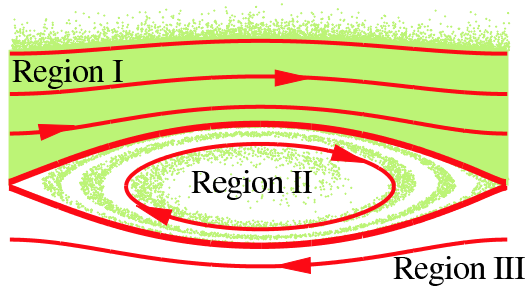}
\caption{Phase space schematic for a monochromatic wave overlaid on the results of a self-consistent particle simulation.  Region I (above the separatrices) consists of the plasma bulk making up the wave; region II contains the trapped particles between the separatrices; region III contains those particles moving too fast to be trapped in the wave.}	\label{fig:portrait}
\end{centering}	
\end{figure}

Here, we include a few important results regarding the pendulum Hamiltonian (\ref{eqn:PendHam}).  The frozen orbits are defined as the level sets $\Ham(p,\theta;\tau) = H$ at a fixed time $\tau$, for which the parameters $\phi_1$ and $\delta\omega$ are constant and the motion is periodic.  A representative phase portrait of the frozen orbits is in Fig.~\ref{fig:portrait}, superposed on a phase-space snapshot taken from a self-consistent particle simulation.  Generically, we see that phase space is divided into three distinct regions, separated by the trajectories joining the hyperbolic fixed points at 
$\theta = \pm\pi,\, p=-\delta\omega$, for which $H=2\phi_1$.  These separatrices separate the rotation motion of regions I and III, for which $H > 2\phi_1$, from the libration about the stable fixed point at 
$\theta = 0$ in region II, where $H<2\phi_1$.  Associated with these frozen orbits, there exists a canonical transformation to action-angle coordinates 
$\Ham(p,\theta;\tau) \rightarrow \Ham(\action,\Psi;\tau)$, with the action proportional to the phase-space area of the frozen orbit:
\beu
  \action(H; \tau) \equiv \frac{1}{2\pi} \oint \! d\theta\; p(\theta,H;\tau).
\eeu
Using (\ref{eqn:PendHam}), the action is given by the the following integral along the frozen orbit:
\be
  \action(\kappa;\tau) = \frac{\kappa\act_s}{4} \oint \! d\theta \;
  	\sqrt{1 - (1/\kappa^2)\sin^2\theta/2}  - \delta\omega(\tau),		\label{eqn:J_int}
\ee
where we have defined the scaled energy $\kappa$ and the separatrix action $\act_s$ as
\begin{align}
  \kappa &\equiv \frac{H}{2\phi_1}	& \act_s &\equiv \frac{4}{\pi}\sqrt{\phi_1}.
  	\label{eqn:Jdefs}
\end{align}
As is well-known, the integral in (\ref{eqn:J_int}) can be evaluated in terms of complete elliptic integrals.  Furthermore, since the orbits change their topology at the separatrix, the action coordinate is discontinuous there.  While no actual particle orbits are singular, this discontinuity does complicate notation; for this reason, we find it convenient to introduce the ``frequency-shifted action'' $\act(\kappa)$ that is 
affinely related to $\action$, and defined according to the phase space region in which the trajectory lives.  Taking the integral in (\ref{eqn:J_int}), we define
\bse		\label{eqn:action}
\begin{align}
  \left\lvert \kappa \right\rvert \ge 1&:& \act(\kappa) &\equiv \action(\kappa;\tau) + \delta\omega(\tau) 
  	= \act_s \kappa\, \mathscr{E}(1/\kappa) \\
  \left\lvert \kappa \right\rvert < 1&:& \act(\kappa) &\equiv
  	\tfrac{1}{2}\left[\action(\kappa;\tau) + \delta\omega(\tau)\right] \nonumber \vphantom{\sum^-} \\
  && &= \act_s \left[\mathscr{E}(\kappa) + (\kappa^2-1)\mathscr{K}(\kappa)\right].
\end{align}
\ese
Here, the complete elliptic integrals of the first and second kind, $\mathscr{K}(\kappa)$ and $\mathscr{E}(\kappa)$, respectively, are defined in the standard way:
\begin{align*}
  \mathscr{K}(\kappa) &\equiv \!\!\int\limits_0^{\pi/2} \!\! \frac{d\alpha}{\sqrt{1 - \kappa^2\sin^2\!\alpha}} &
  \mathscr{E}(\kappa) &\equiv \!\!\int\limits_0^{\pi/2} \!\! d\alpha\, {\sqrt{1 - \kappa^2\sin^2\!\alpha}}.
\end{align*}
Furthermore, the nonlinear period of the pendulum is simply calculated using the Hamiltonian relations
\beu
  \mathcal{T}(\kappa) = \frac{2\pi}{\omega(\kappa)} = 2\pi \frac{\partial\action}{\partial\Ham} = 
  	2\pi \frac{\partial\action}{\partial\kappa}\frac{\partial\kappa}{\partial\Ham}.	
\eeu
From the definitions (\ref{eqn:Jdefs}) and (\ref{eqn:action}), we have
\bse		\label{eqn:period}
\begin{align}
  \left\lvert \kappa \right\rvert \ge 1 : \hspace{.1 in} \mathcal{T}(\kappa) &= 
  	\frac{2}{\kappa\sqrt{\phi_1}}\mathscr{K}(1/\kappa) \\
  \left\lvert \kappa \right\rvert < 1 : \hspace{.1 in} \mathcal{T}(\kappa) &= 
  	\frac{4}{\sqrt{\phi_1}}\mathscr{K}(\kappa).
\end{align}
\ese

Although the transformation $(p,\theta)\rightarrow (\action,\Psi)$ is canonical, we note that the variables $(\act,\Psi)$ do not form a canonical pair due to the scaling at the separatrix.  Rather than use these variables to calculate phase space averages, more straightforward analysis is obtained using the scaled energy $\kappa$ and the time $\tau$.  Thus, we conclude by relating the coordinates $(\kappa,\tau)$ to $(p,\theta)$.  Using the defintions (\ref{eqn:pqDefs}) and (\ref{eqn:Jdefs}), we have
\be
  p + \delta\omega(\tau) = \frac{d\theta}{d\tau} = 2\kappa\sqrt{\phi_1}\sqrt{1 - (1/\kappa^2)\sin^2(\theta/2)}.
		\label{eqn:p_theta}
\ee
Rewriting this expression, we find
\bse		\label{eqn:DphiDt}
\begin{align}
  \left\lvert\kappa\right\rvert &\ge 1: & \frac{d(\theta/2)}{\sqrt{1 - (1/\kappa^2)\sin^2(\phi/2)}} 
	&= \kappa\sqrt{\phi_1}\, d\tau \\
  \left\lvert\kappa\right\rvert &< 1: &  \frac{d\alpha}{\sqrt{1 - \kappa^2\sin^2\alpha}} &= \sqrt{\phi_1}\, d\tau
\end{align}
\ese
with $\sin(\theta/2) \equiv \kappa\sin\alpha$.  We take the indefinite integral of (\ref{eqn:DphiDt}), obtaining
\bse	\label{eqn:HamJacobi}
\begin{align}
  \left\lvert\kappa\right\rvert \ge 1: \, \vphantom{t_{\sum_j}} \cos(\theta/2) &= 
  	\text{cn}\!\left(1/\kappa, \kappa\sqrt{\phi_1}\tau \right) \vphantom{\lim_{-}} \\
  \left\lvert\kappa\right\rvert < 1: \, \cos(\theta/2) &= \text{dn}\!\left(\kappa, \sqrt{\phi_1}\tau \right),
\end{align}
\ese
where, without loss of generality, we have set the origin of time to zero, and the functions $\text{cn}(\kappa,x)$ and $\text{dn}(\kappa,x)$ are the Jacobian elliptic functions defined in the usual manner via the inverse of the incomplete elliptic integral of the first kind:
\beu
\begin{split}
  \text{for }x(\kappa, y) \equiv \int\limits_0^y \! dz\; 
  	\frac{1}{\sqrt{1-\kappa^2\sin^2 z}}: \phantom{hm}\\
  \cos y \equiv \text{cn}(\kappa,x) \equiv 
  	\frac{1}{\kappa}\sqrt{\kappa^2 - 1 + \text{dn}^2(\kappa,x)}.
\end{split}
\eeu
Finally, differentiating (\ref{eqn:HamJacobi}) and using (\ref{eqn:p_theta}) yields
\bse	\label{eqn:pHamJacobi}
\begin{align}
  \left\lvert\kappa\right\rvert \ge 1:  \vphantom{t_{\sum_j}}\, p &= 
	2\kappa\sqrt{\phi_1} \, \text{dn}\!\left(1/\kappa, \kappa\sqrt{\phi_1}\tau \right) - \delta\omega(\tau) 
	 \vphantom{\lim_{-}} \\
  \left\lvert\kappa\right\rvert < 1: \, p &= 
  	2\kappa\sqrt{\phi_1} \, \text{cn}\!\left(\kappa, \sqrt{\phi_1}\tau \right) - \delta\omega(\tau).
\end{align}
\ese

\section{The nonlinear distribution function action \emph{ansatz}}		\label{section:distribution}

In this section, we introduce our BGK-type distribution function \emph{ansatz} that naturally arises in weakly driven plasmas.  This results in a dramatic reduction in the number of degrees of freedom from the Vlasov equation (\ref{eqn:Vlasov}), and therefore can be used as the foundation of a simplified, fluid-like model.

We first motivate our assumed functional form for $f$, showing how simple physical arguments based on particle phase mixing in the slowly changing electrostatic field governed by Vlasov evolution lead to a very simple distribution: gaussian in the canonical action $\action$ with slowly evolving mean and fixed variance.  We include an example from a slowly driven, self-consistent particle simulation that naturally evolves in this way, and compare it to the more familiar velocity-space distributions.  Finally, we give the distribution function in an explicit form from which phase-space averages can be computed. 

\subsection{Phase-mixing in the slowly-growing wave}

The central assumption for our distribution function action \emph{ansatz} is that the Langmuir wave amplitude and frequency are slowly evolving, meaning that the parameters $\phi_1(\tau)$ and $\delta\omega(\tau)$ of the pendulum Hamiltonian (\ref{eqn:PendHam}) do not vary appreciably over the period $\mathcal{T}$.  For the nearly all electrons, i.e., all except the exponentially few in a narrow range about the separatrix, the condition of ``slowness'' can be written as
\begin{align}
  \frac{1}{\sqrt{\phi_1}}\frac{d}{d\tau}\left\lvert \ln\phi_1 \right\rvert &\sim 
  	\frac{1}{\sqrt{\phi_1}}\frac{d}{d\tau} \left\lvert \ln\delta\omega \right\rvert \sim \epsilon, &
	\epsilon \ll 1.	\label{eqn:slow}
\end{align}
As previously noted, while these conditions are reminiscent of adiabatic motion, we do not explicitly invoke adiabaticity; rather, we use the slowness condition (\ref{eqn:slow}) to justify our assumption that the distribution function remains essentially uniform in the canonical angle $\Psi$ throughout its evolution.  Far from the separatrices, it is clear that the slowness conditions (\ref{eqn:slow}) imply that the electrons make many oscillations before the parameters of the wave significantly change, so that a set of these particles that is initially uniform in canonical angle remains so under evolution by (\ref{eqn:PendHam}).  As particles approach the separatrix where $\kappa \rightarrow 1$ and the nonlinear period diverges $\sim \! \ln\left\lvert 1 - \kappa^2 \right\rvert$, such a simple argument breaks down.  In this case, we invoke the results of phase evolution by Cary and Skodje \cite{CarySkodje:1988, CarySkodje:1989} and Elskens and Escande \cite{ElskensEscande:1991}, obtained by analyzing the near-separatrix motion in slowly evolving systems.

For most of the particles, the canonical angle is mapped smoothly through the separatrix.  In a naive picture, the strip of particles in region I with the same action $J$ and spread over $0 \le \Psi < 2\pi$ is mapped across the separatrix to the strip from $0 \le \Psi < \pi$ that then rotates in region II.  As shown in 
 \cite{CarySkodje:1988, CarySkodje:1989, ElskensEscande:1991}, this picture is essentially correct to $O(\epsilon)$, excluding the exponentially few $O \! \left(e^{-1/\epsilon}/\epsilon \right)$ particles that pass very close to the hyperbolic fixed point.  Since these particles can spend an arbitrarily long time tracing the stable manifold, they lead to long, diffuse phase-space tendrils.  Neglecting these particles, to each action is associated a strip of particles that is mapped to one-half the canonical angle upon crossing the separatrix.  This proceeds with each successive action strip, with each one displaced from the next by a relative phase in the canonical angle.  The relative phase between increasing action strips increases up to $2\pi$, for which the action has increased by $O(\epsilon)$ (see 
\cite{ElskensEscande:1991} for a detailed discussion).  In this way, we argue that the distribution remains phase mixed to $O(\epsilon)$ even when crossing the separatrix, provided only that the slowness conditions (\ref{eqn:slow}) are met.

Assuming that the electrostatic wave amplitude and phase velocity are slowly varying, the previous arguments indicate that the distribution function remains uniform in the canonical angle throughout the range $0 \le \Psi < 2\pi$, resulting in the following simplification:
\be
  f(\action,\Psi; \tau) \rightarrow  \tfrac{1}{2\pi}f(\action; \tau).
\ee
To make further progress, we also assume that the distribution $f(\action;\tau)$ is well-represented by its first two moments in canonical action.  This is trivially true if $\phi_1 = 0$, for which $v \propto \action$ and a Maxwellian plasma is gaussian in action.  In the general case, this assumption is similar in spirit to a warm-fluid theory (see, e.g., \cite{Newcomb:1982, Shadwick_et_al:2004}), for which asymptotic solutions are obtained by systematically neglecting the (presumably small) third- and higher-order moments in $v$.  By using moments in $\action$, however, our model will self-consistently include the effects of trapped particles on the Langmuir wave, as do the models of Holloway, Dorning and Buchanan \cite{HollowayDorning:1991, BuchananDorning:1995}; on the other hand, unlike the theories of 
\cite{Newcomb:1982, Shadwick_et_al:2004}, ours is non-relativistic.  Retaining only the first two moments in the canonical action is equivalent to prescribing $f$ to be gaussian in $\action$:
\be
  f(\action;\tau) = \frac{1}{\sigma(\tau)\sqrt{2\pi}} \exp\!\left\{\tfrac{\left[\action - \bar{\action}(\tau)\right]^2}
  	{2\sigma(\tau)^2}\right\}.	\label{eqn:firstGaussian}
\ee
We can simplify (\ref{eqn:firstGaussian}) in the following manner.  The Vlasov equation (\ref{eqn:Vlasov}) permits an infinite number of conservation laws (Casimirs), of which the entropy is one of particular physical significance:
\be
  \frac{d}{d\tau} \int \! d\action d\Psi \; f(\action,\Psi;\tau) \ln f(\action,\Psi;\tau) = 0.	\label{eqn:CasimirS}
\ee
Using the gaussian-in-action distribution (\ref{eqn:firstGaussian}), the conservation of entropy (\ref{eqn:CasimirS}) implies that
\be
  \frac{d}{d\tau} \ln \sigma(\tau) = 0,	\label{eqn:entropy}
\ee
i.e., that the gaussian width is fixed throughout the evolution.  Some care must be taken to apply this result across the separatrix where the action $\action$ is not continuous; note that this is an issue of coordinates, not dynamics.  Since the coordinate $\action$ doubles across the separatrix, the width in $\action$ similarly doubles.  To be more precise, the phase space region $\left\{[\action_1, \action_1+\delta\action], \; [0, 2\pi) \right\}$ just above (or below) the separatrix corresponds to the equal area region between the separatrices over the intervals 
$\left\{[2\action_1, 2(\action_1+\delta\action)], \;[0, \pi) \right\}$.  
Thus, in terms of the scaled actions $\act(\kappa)$ defined in (\ref{eqn:action}), the width $\sigma$ is the same in all regions.  Using the results (\ref{eqn:firstGaussian}) and (\ref{eqn:entropy}), we arrive at the following form for the distribution function
\be
  f(\kappa;\tau) = \frac{1}{2\pi}\frac{1}{\sigma\sqrt{2\pi}}\exp\!\left\{ -\tfrac{1}{2\sigma^2}\left[\act(\kappa) - 
  	\bar{\act}(\tau)\right]^2\right\}.	\label{eqn:f_kappa}
\ee
Note that as defined the initial width is a measure of the temperature such that $\sigma=k\lambda_D$.

\begin{figure}
\begin{centering}
\includegraphics[scale=0.76]{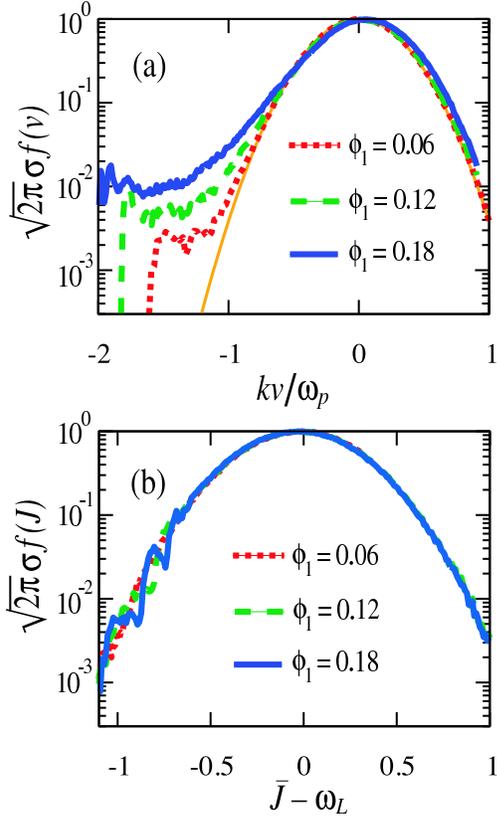}
\caption{Wavelength averaged distribution functions for three values of the electrostatic potential $\phi_1$ using $k_2\lambda_D = 0.3$.  In velocity-space, (a) shows the flattening of $f(v)$ about the phase velocity $k v/\wp \approx 1.16$.  Meanwhile, (b) demonstrates that the distribution in canonical action remains nearly gaussian throughout the excitation of the plasma wave.}	\label{fig:distribution}
\end{centering}	
\end{figure}

To illustrate the distributions associated with (\ref{eqn:f_kappa}), we have performed a number of single-wavelength particle simulations, described in greater detail in Sec.~\ref{section:simulation}, that solve the periodic Vlasov-Poisson system.  We include representative results in Fig.~\ref{fig:distribution}, obtained with a drive potential $V = 0.01$ and initial $\sigma=0.3$.  In Fig.~\ref{fig:distribution}(a) we see the characteristic flattening of $f(v)$ near the phase velocity that is associated with particle trapping.  For the same values of $\phi_1$, Fig.~\ref{fig:distribution}(b) demonstrates that the distribution in $\action$ (integrated over $\Psi$) remains nearly gaussian in the canonical action.  Furthermore, except for the slight oscillations near the resonance region, $f(\action)$ has a constant variance $\sigma^2$ and a slightly increasing mean (from $\omega_L$).  Note that Fig.~\ref{fig:distribution} uses the numerically calculated $J$ from the exact simulation potential, which we find only differs from the pendulum results very near the separatrix.

\subsection{Phase space averages}
Here, we combine the distribution function (\ref{eqn:f_kappa}) with our assumptions on the slowly-varying nature of $\bar{\act}$ to arrive at a convenient expression of phase space averages in terms of the variables $\kappa$ and $\tau$.  First, we note that for fixed parameters $\phi_1$, $\delta\omega$, there exists a canonical transformation of $\mathcal{H} : (p,\theta) \rightarrow (H,\tau)$ for which the evolution parameter is the coordinate $\theta$.  Since the transformation is canonical, the Jacobian is unity and we have the following relation between the integration measures:
\be
  dp\, d\theta = dH\, d\tau = \frac{dH}{d\kappa} d\kappa\, d\tau = 4\phi_1 \kappa \; d\kappa\, d\tau.
\ee
Thus, phase space averages can be written as
\begin{align}
  \big\langle\mathcal{X}(p,\theta)\big\rangle &\equiv
  	\int\limits_{-\infty}^\infty \!\! dp \int\limits_{-\pi}^{\pi} \!\! d\theta \; f(p,\theta;\tau) \mathcal{X}(p,\theta) 
	\nonumber \\
  &= \int\limits_{-\infty}^\infty \!\! d\kappa \; 4\phi_1\kappa 
	\int\limits_0^\mathcal{T} \! d\tau \; f(\kappa;\tau) \mathcal{X}(\kappa,\tau).
		\label{eqn:FirstAve}
\end{align}
Assuming that the average action does not change significantly during one period, i.e., that the conditions (\ref{eqn:slow}) are met, we can take the nearly constant distribution $f(\kappa;\tau)$ through the integral over $\tau$.  Hence, in subsequent calculations we suppress the dependence of $f$ on $\tau$.  To further simplify notation and integration limits, we assign the following expressions for the distribution in $\kappa$ appropriate to the various phase space regions:
\bse		\label{eqn:fDefs}
\begin{align}
  f_\text{I}(\kappa) &\equiv \frac{\act_s \mathscr{K}(1/\kappa)}{\sigma\sqrt{2\pi}}
  	\exp\!\left\{-\tfrac{1}{2\sigma^2}\left[ \act(\kappa) - \bar{\act}\right]^2\right\}  \\
  f_\text{III}(\kappa) &\equiv \frac{\act_s \mathscr{K}(1/\kappa)}{\sigma\sqrt{2\pi}}
  	\exp\!\left\{-\tfrac{1}{2\sigma^2}\left[ \act(\kappa) + \bar{\act}\right]^2\right\} \\
  f_\text{II}^\pm(\kappa) &\equiv \frac{\act_s \kappa\mathscr{K}(\kappa)}{\sigma\sqrt{2\pi}}
  	\exp\!\left\{-\tfrac{1}{2\sigma^2}\left[\act(\kappa) \pm \bar{\act} \right]^2\right\}.
\end{align}
\ese
The definitions (\ref{eqn:fDefs}) arise by absorbing the factor of $4\phi_1\kappa$ into the $f$ of 
(\ref{eqn:f_kappa}), appropriately changing the sign of $\kappa$, and multiplying by the period $\mathcal{T}$ from (\ref{eqn:period}); these conventions imply that
\beu
  \int\limits_1^\infty \! d\kappa \; \big[ f_\text{I}(\kappa) + f_\text{III}(\kappa)\big] + \int\limits_0^1 \!
  	d\kappa \; \big[ f_\text{II}^-(\kappa) + f_\text{II}^+(\kappa)\big] = 1.
\eeu
Using the definitions (\ref{eqn:fDefs}) in (\ref{eqn:FirstAve}), and remembering the chosen sign conventions for $\kappa$ and the scaling by $\mathcal{T}$, phase space averages are computed with the integral expression
\be
\begin{split}
  \big\langle \mathcal{X} \big\rangle &= \int\limits_1^\infty \! d\kappa \int\limits_0^\mathcal{T} \! 
  	d\tau  \left[ f_\text{I}(\kappa) \tfrac{\mathcal{X}(\kappa,\tau)}{\mathcal{T}(\kappa)} + 
	f_\text{III}(\kappa)\tfrac{\mathcal{X}(-\kappa,\tau)}{\mathcal{T}(\kappa)} \right] \\
  &\phantom{i} + \int\limits_0^1 \! d\kappa \int\limits_0^\mathcal{T} \! d\tau \left[ 
  	f_\text{II}^{-}(\kappa)\tfrac{\mathcal{X}(\kappa,\tau)}{\mathcal{T}(\kappa)} + 
	f_\text{II}^{+}(\kappa)\tfrac{\mathcal{X}(-\kappa,\tau)}{\mathcal{T}(\kappa)}\right].
		\label{eqn:average}
\end{split}
\ee

\section{Nonlinear self-consistency: parameterizing the distribution}	\label{section:BGK}

Now that we have presented the distribution function \emph{ansatz} and developed the necessary pendulum machinery, we turn to calculating expressions for the mean action $\bar{\act}$ and the frequency shift $\delta\omega$.  We do this by imposing the constraints of self-consistency: the assumed distribution function must also satisfy Maxwell's equations.  After obtaining $\bar{\act}$ and $\delta\omega$ in terms of integrals over the modulus of the elliptic functions, we will derive small-amplitude, analytic expressions suitable to comparisons with previous published results.  In the next section we compare these theoretical results to numerical examples from a self-consistent particle code.

\subsection{Self-consistency and Maxwell's equations}

First, our nonlinear distribution function must give rise to a charge separation commensurate with the electrostatic potential $\phi(z,t)$.  This is summarized by the Poisson equation:
\be
  \frac{\partial^2\Phi}{\partial z^2} =  4\pi e \! \int \! dv\; f(v,z;t) - 4\pi e n_i(z).		\label{eqn:firstPoisson}
\ee
Fourier transforming (\ref{eqn:firstPoisson}), assuming the ions are stationary $[n_i(z,t)=n_0]$, and using the dimensionless Fourier expansion of the potential (\ref{eqn:FourPhi}), the Poisson equation for the lowest harmonic of the potential can be written as
\be
  0 = \left[ 1 + \frac{2}{\phi_1} \big\langle \cos\theta(\kappa,\tau)\big\rangle \right]\phi_1 
  	\equiv \Ez\!\left(\bar{\act}; \phi_1\right) \phi_1,	\label{eqn:NLdiel}
\ee
where we have introduced the ``nonlinear dielectric'' $\Ez\!\left(\bar{\act}; \phi_1\right)$.  From 
(\ref{eqn:NLdiel}), we see that for our assumed distribution to support a nontrivial potential $\phi_1$, the nonlinear dielectric must vanish.  This is an implicit relation for the mean action $\bar{\act}$ in terms of the potential amplitude, i.e., given a potential $\phi_1$, the requirement $\Ez\!\left(\bar{\act}; \phi_1\right) = 0$ establishes the appropriate value of $\bar{\act}$.  To determine the nonlinear dielectric, we use our assumed distribution function to calculate the phase space average of $\cos\theta(\kappa,\tau)$ as indicated in (\ref{eqn:NLdiel}).  We define the variables 
$x\equiv \kappa\sqrt{\phi_1}\tau$, $y\equiv \sqrt{\phi_1}\tau$, and use the pendulum identity 
(\ref{eqn:HamJacobi}) and the phase space averaging (\ref{eqn:average}), to arrive at
\begin{align*}
\begin{split}
  \big\langle \cos\theta \big\rangle &= \int\limits_1^\infty \!\! d\kappa \; 
  	\big[ f_\text{I}(\kappa) + f_\text{II}(\kappa) \big] \!\!\! \int\limits_0^{\mathscr{K}(1/\kappa)} \!\!\!\!\! dx \; 
	\frac{2\text{cn}^2\!\left(1/\kappa, x \right) - 1}{\mathscr{K}(1/\kappa)} \\
 &\phantom{=} + \int\limits_0^1 \!\! d\kappa \; 
  	\left[ f_\text{II}^{-}(\kappa) + f_\text{II}^{+}(\kappa) \right] \!\! \int\limits_0^{\mathscr{K}(\kappa)} \!\!\! 
	dy \; \frac{2\text{dn}^2\!\left(\kappa, y \right) - 1}{\mathscr{K}(\kappa)}.
\end{split}
\end{align*}
The integrals over $x$ and $y$ can be taken analytically; using the integral tables of Gradshteyn and Ryzhik \cite{gradshteyn_ryzhik:80}, the nonlinear dielectric is given by
\be
\begin{split}
  \Ez\!\left(\bar{\act},\phi_1\right) &= 1 + \frac{2}{\phi_1} \int\limits_1^\infty \! d\kappa\;
	\big[f_\text{I}(\kappa) + f_{\text{III}}(\kappa)\big] \\
  &\phantom{=| 1 + \frac{2}{\phi_1}\int\limits_1^\infty \! d\kappa} \,
	\times\!\left[\frac{2\kappa^2\mathscr{E}(1/\kappa) }{\mathscr{K}(1/\kappa)} + 1 - 2\kappa^2 \right] \\
  &\, + \frac{2}{\phi_1} \int\limits_0^1 \! d\kappa \,
  	\big[f_\text{II}^+(\kappa) + f_{\text{II}}^{-}(\kappa)\big] \!
	\left[\frac{2\mathscr{E}(\kappa)}{\mathscr{K}(\kappa)} - 1 \right]\!.	\label{eqn:dielectric1}
\end{split}
\ee
Requiring (\ref{eqn:dielectric1}) to vanish gives an implicit relationship between $\phi_1$ and $\bar{\act}$; to obtain an explicit expression for the mean frequency-shifted action in terms of the potential, we expand the nonlinear dielectric for small changes in $\bar{\act}$.  For an initially Maxwellian plasma with zero electrostatic field, the mean action in the moving frame is equal to the linear frequency $\omega_L$, and we have the expansion
\be
  0 = \Ez(\omega_L,\phi_1) + \left(\bar{\act} - \omega_L\right) \left. \frac{\partial}{\partial\bar{\act}} 
	\, \Ez\!\left(\bar{\act},\phi_1\right) \right\rvert_{\bar{\act}=\omega_L} \!\! + \cdots .
		\label{eqn:TaylorDiel}
\ee
Rewriting the Taylor expansion (\ref{eqn:TaylorDiel}), the change in the mean action in terms of potential amplitude is
\be
  \bar{\act} - \omega_L \approx - \frac{\Ez(\omega_L,\phi_1)}{\tfrac{\partial}{\partial\omega_L}\,
  	\Ez(\omega_L,\phi_1)}.		\label{eqn:MeanAct}
\ee
Equation (\ref{eqn:MeanAct}) determines the mean frequency-shifted action $\bar{\act}$ required to support a potential of amplitude $\phi_1$.  In previous works (e.g., \cite{MoralesONeil:1972, RoseRussell:2001}), expressions similar to (\ref{eqn:MeanAct}) were interpreted as the frequency shift of the wave.  We will see that in the small $\phi_1$ limit, the change in $\bar{\act}$ equals the change in the frequency.  This is because the particle action $\action$ is essentially constant in this case, so that $\act \propto p$ implies that a change in $\act$ is due to a decrease in the phase velocity of the wave at fixed $k$ [i.e., given by $\delta\omega(\tau)$].  As the wave amplitude becomes appreciable, however, the plasma bulk becomes excited and the increase in canonical action $\action$ can no longer be neglected, so that $\delta\omega(\tau) < \bar{\act}(\tau)-\omega_L$ and (\ref{eqn:MeanAct}) is numerically larger than the (generically negative) frequency shift.

To complete the characterization of the action distribution function, we calculate an expression for the frequency shift as a function of $\phi_1$ and the mean action $\bar{\act}$.  To determine $\delta\omega$, we use the fact that the plasma tends to set up a return current to erase any long-range electric fields.  This is related to the fact that in 1-D with immobile ions, the electrons gain no net momentum.  To make this more explicit, we consider the 1-D Coulomb gauge condition: 
$\boldsymbol{\nabla}\cdot\boldsymbol{A} = 
\tfrac{\partial}{\partial z}\big(\unitvec{z} \cdot \boldsymbol{A}\big) = 0$, implying that the vector potential can be taken to be purely transverse, i.e., $\boldsymbol{A} \cdot \unitvec{z} = 0$.  In this case, the longitudinal component of the Ampere-Maxwell equation is given by
\be
  \frac{1}{c}\frac{\partial^2}{\partial t\partial z}\Phi(z,t) 
  	= 4\pi e \! \int \!\! dv \; f(v,z;t)\, v.	
\ee
Integrating 
over one period in $z$, we have
\be
  \frac{\partial}{\partial t} \left[\Phi\left(\tfrac{\pi}{k},t\right) - \Phi\left(-\tfrac{\pi}{k},t\right)\right] = 4\pi e \! 
  	\int\limits_{-\infty}^\infty \!\! dv \int\limits_{-\pi}^\pi \! dz \; f \, v.	\label{eqn:AveAmpMax}
\ee
Since we assume the potential $\Phi(z,t)$ to be slowly varying, the expression (\ref{eqn:AveAmpMax}) approximately vanishes.  Note that this is exact in the limit of a time-independent, nonlinear mode (similar to BGK), and expresses the fact that the plasma electrons carry no net momentum \cite{McKinstrieYu:1991}.  Using the fact that the velocity 
$v \propto (p - \omega_L)$, we have
\be
  \big\langle p(\kappa,\tau) \big\rangle - \omega_L = 0.
\ee
To evaluate this phase space average, we use the pendulum formula (\ref{eqn:pHamJacobi}), which gives the momentum in terms of Jacobi elliptic functions and the frequency shift $\delta\omega(\tau)$.  Since the averaged velocity of the trapped particles in the moving frame is zero [the integral of $\text{cn}(\kappa,x)$ from 0 to $4\mathscr{K}(\kappa)$ vanishes], we have
\beu
\begin{split}
  \omega_L + \delta\omega(\tau) &= \int\limits_1^\infty \! d\kappa\; \frac{f_\text{I}(\kappa) - 
  	f_\text{III}(\kappa)}{\mathscr{K}(1/\kappa)} \!\!
  	\int\limits_0^{\mathscr{K}(1/\kappa)} \!\!\!  dx \; \text{dn}(1/\kappa, x).
\end{split}
\eeu
We take the integral over $x \equiv \kappa\sqrt{\phi_1}\tau$ using Gradshteyn and Ryzhik \cite{gradshteyn_ryzhik:80}, so that
\be
\begin{split}
  \delta\omega(\tau) &= \int\limits_1^\infty \! d\kappa\;
  	\frac{4\phi_1\kappa}{\sqrt{2\pi}\sigma}\,
	e^{-\tfrac{1}{2\sigma^2}\left[\act_s\kappa\mathscr{E}(1/\kappa) - \bar{\act}\right]^2}  \\
  &\hphantom{=|}  - \int\limits_1^\infty \! d\kappa\; \frac{4\phi_1\kappa}{\sqrt{2\pi}\sigma}
  	e^{-\tfrac{1}{2\sigma^2}\left[\act_s\kappa\mathscr{E}(1/\kappa) + \bar{\act}\right]^2} - \omega_L.		\label{eqn:FreqShift}
\end{split}
\ee

\subsection{The mean action $\bar{\act}$ and frequency shift $\delta\omega$: linear and small amplitude limits}

In this section we study the linear and small amplitude limits of $\bar{\act}$ and $\delta\omega$, and compare them to previous results.  First, we present the linear limit of (\ref{eqn:MeanAct}), for which $\phi_1 \rightarrow 0$.  In this limit, the mean action is that corresponding to the phase velocity of the infinitesimal wave, so that $\bar{\act} \rightarrow \omega_L$.  Since the denominator in (\ref{eqn:MeanAct}) is well-behaved, the linear limit is characterized by
\be
  \lim_{\phi_1\rightarrow 0}\Ez (\omega_L,\phi_1) = 0.	\label{eqn:limDisp}
\ee
For clarity, we reserve the cumbersome calculations used to evaluate (\ref{eqn:limDisp}) for the Appendix.  In the Appendix, we show that our assumed distribution gives a concrete prescription for the usual pole occurring when the particle velocity matches that of the wave phase velocity.  Denoting the principal value by $\mathscr{P}$, from (\ref{eqn:limDispP}) we have
\be
  \lim_{\phi_1\rightarrow 0} \Ez(\omega_L,\phi_1)  = 1 + \frac{1}{\sigma^2} + 
	\frac{\omega_L/\sigma}{\sigma^2\sqrt{2\pi}}\; \mathscr{P} \!\!
	\int\limits_{-\infty}^\infty \!\! dx\; \frac{e^{-x^2/2}}{x - \omega_L/\sigma} .	\label{eqn:DerivedDisp}
\ee
Setting the linear dielectric (\ref{eqn:DerivedDisp}) to zero yields the plasma dispersion relation as found by Vlasov \cite{Vlasov:1945}, resulting in a purely real natural frequency.  Physically, this lack of linear Landau damping arises because we have assumed that the distribution is completely phase mixed.  As shown by O'Neil \cite{ONeil:1965}, such phase mixing causes linear Landau damping to be a transient effect that itself damps away on the bounce time scale $\sim 1/(\wp\sqrt{\phi_1})$.  Although the bounce time diverges as $\phi_1 \rightarrow 0$, we maintain that our analysis and the dispersion relation 
(\ref{eqn:DerivedDisp}) applies to finite amplitude waves after several bounce periods have passed.  In this case, the distribution is nearly uniform in canonical angle and Landau damping has been ``washed out''.  The nonlinear fate of such Langmuir waves is generally considered to be a BGK wave; as discussed in \cite{HollowayDorning:1991} and \cite{BuchananDorning:1995}, the dispersion relation of small amplitude, sinusoidal BGK waves is that of Vlasov, and is identical to (\ref{eqn:DerivedDisp}) derived here.

For $\phi_1$ small but not infinitesimal, we Taylor expand the dielectric (\ref{eqn:dielectric1}) in the Appendix, yielding
\begin{align}
  \Ez(\omega_L,\phi_1) &\approx 1.089\sqrt{\phi_1} 
  	\left( \frac{\omega_L^2}{\sigma^2} - 1 \right)
	\frac{e^{-\omega_L^2/2\sigma^2}}{\sqrt{2\pi}\sigma^3}.	\label{eqn:SmDielectric}
\end{align}
A similar expression for the small amplitude dielectric has been derived by a number of authors, although there is some variation in the $O(1)$ coefficient replacing our 1.089.  Interestingly, our coefficient is precisely that of Dewar \cite{Dewar:1972}, who calculated the frequency shift assuming a small but finite sinusoidal wave that is adiabatically excited; other calculations in a similar regime have obtained values of 1.41 (Manheimer and Flynn \cite{ManheimerFlynn:1971}), 1.76 (Rose and Russell \cite{RoseRussell:2001}) and 1.60 (Barnes \cite{Barnes:2004}).  It should be noted that these differ slightly from the coefficient of 1.63 calculated by Morales and O'Neil \cite{MoralesONeil:1972} and separately by Dewar \cite{Dewar:1972} for an instantaneously excited wave (i.e., the initial value problem).  The majority of these authors then use this to determine the nonlinear frequency shift using an expression similar to (\ref{eqn:MeanAct}); we will see that this yields reasonable results provided that $\phi_1 \lesssim \sigma^2$.

To complete the small amplitude analysis of the change in the average action (\ref{eqn:MeanAct}), we differentiate the linear dispersion relation (\ref{eqn:DerivedDisp}), obtaining $\tfrac{\partial}{\partial\omega_L}\Ez(\omega_L,\phi_1)$ in the small amplitude limit.  Using 
(\ref{eqn:MeanAct}) and (\ref{eqn:SmDielectric}), we find that the change in the mean action is expressed for small values of $\phi_1$ by
\begin{align}
  \bar{\act}(\tau) - \omega_L &\approx - 1.089\,\omega_L\sqrt{\phi_1}
  	\frac{\left(\tfrac{\omega_L^2}{\sigma^2} - 1\right)e^{-\omega_L^2/2\sigma^2}}
  	{\left(\omega_L^2 - 1 - \sigma^2\right)\sqrt{2\pi}\sigma} 	\nonumber \\
  &= -1.089\sqrt{\phi_1}\frac{\omega_L\sigma^2}{\omega_L^2 - 1 - \sigma^2}
	f''(\action)\Big\rvert_{\begin{subarray}{l} \action=0 \\ \bar{\action}=\omega_L
					\end{subarray}} \!,	\label{eqn:SmMean}
\end{align}
with the distribution $f(\action)$ given by (\ref{eqn:firstGaussian}).

\begin{figure*}
\begin{centering}
\includegraphics[scale=.56]{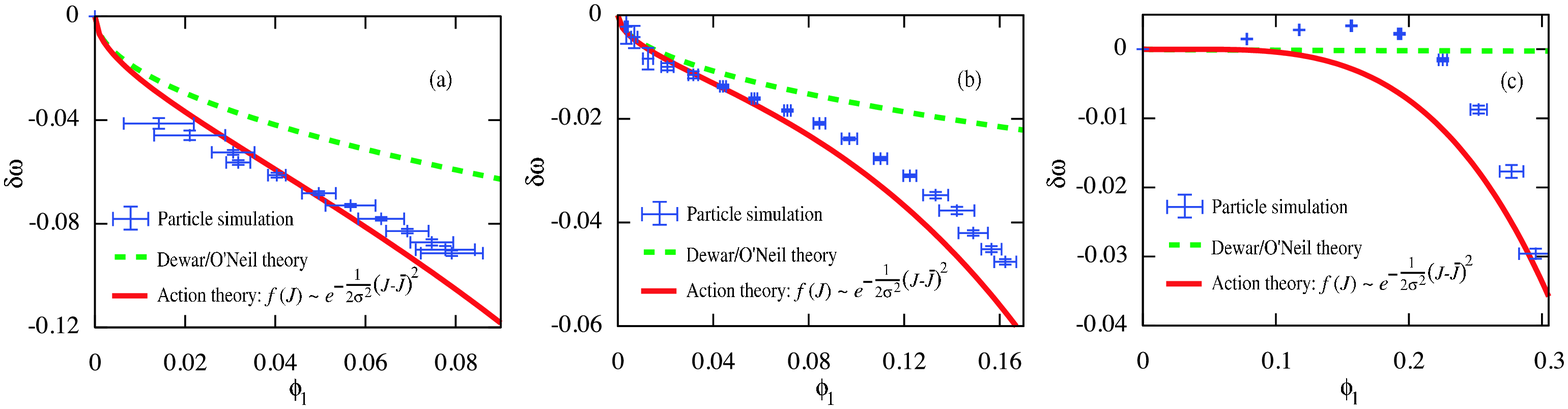}
\caption{Frequency shift for three different temperatures: $k\lambda_D=0.4$ (a), $k\lambda_D=0.3$ (b), and $k\lambda_D=0.2$ (c).  The  points are from simulation results, with error bars indicating two standard deviations in the measured data.  We see that the theoretical value of $\delta\omega$ from the action distribution (solid line) agrees quite well with the simulation results, which are only closely represented by the Dewar/O'Neil theory (\ref{eqn:SmMean}) for small values of $\phi_1$.  To calculate our theoretical frequency shift, we evaluate the numerical integral (\ref{eqn:FreqShift}) with the mean $\bar{\act}$ evaluated numerically using the approximate expression (\ref{eqn:MeanAct}).}
	\label{fig:frequency}
\end{centering}	
\end{figure*}

To conclude this section, we give the difference between the frequency shift $\delta\omega$ and the shift in action $\bar{\act}-\omega_L$, showing that these differ as the plasma wave amplitude $\phi_1$ grows.  In the Appendix, we see that this difference is given by
\beu
  \delta\omega \approx \big(\bar{\act} - \omega_L\big) - \frac{64}{9\pi}
  	\frac{\omega_L}{\sigma^2} \frac{e^{-\omega_L^2/2\sigma^2}}{\sigma\sqrt{2\pi}} \phi_1^{3/2} + 
	\ldots.
\eeu

\section{Comparison to self-consistent particle simulations}	\label{section:simulation}

In this section, we compare our theoretical results for the properties of the slowly-driven nonlinear Langmuir waves with those obtained from particle simulations.  Before discussing these examples, we make a few comments on the numerical methods.  In these single-wavelength simulations, we numerically solve the equations of motion for the electrons and the electric field, driven by an external force.  For a wavelength with $N$ electrons, the electron with coordinate $\zeta_j \equiv k_2 z_j$ experiences the combined self-consistent electrostatic force and prescribed drive, giving rise to the following equation of motion:
\begin{align}
\begin{split}
  \frac{d^2}{d\tau^2} \zeta_j(\tau) &= \sum_{m =1}^M\frac{1}{N}\sum_{\ell=1}^N\frac{2}{m}
  	\sin\!\big[m\zeta_j(\tau) - m\zeta_\ell(\tau)\big]		\\
  &\phantom{=|} + V(\tau)\cos(\omega_L\tau + \zeta_j) - \En_0(\tau),	\label{eqn:particles}
\end{split}
\end{align}
where we have expanded the electrostatic potential in M harmonics, each of which is a sum over the $N$ macro-particles.  This is a standard technique of the free electron laser community 
\cite{ColsonRide:1980}, although here we have also retained the DC field $\En_0$ 
\cite{Shvets_et_al:1997}, to be calculated using the longitudinal component of the Ampere-Maxwell law:
\be
  \frac{d}{d \tau} \En_0(\tau) = \frac{1}{N}\sum_{\ell=1}^N \frac{d}{d\tau}\zeta_\ell(\tau).	\label{eqn:E0}
\ee
For the examples shown here, we use $N \approx 10^6$ particles and $M \approx 32$ harmonics.  We solve the system (\ref{eqn:particles}-\ref{eqn:E0}) for a given drive potential $V(\tau)$ using a symplectic operator-splitting scheme that is second-order accurate.

To compare the simulation results to our theory, we slowly turn on the ponderomotive drive, ramping the electrostatic field to a chosen amplitude, at which time we slowly turn the drive off.  By taking the Hilbert transform of the potential we obtain the total frequency $(\omega_L + \delta\omega)$, from which we extract the frequency shift for a given amplitude $\phi_1$.  These results are shown in 
Fig.~\ref{fig:frequency}, where we compare the frequency shift extracted from simulation to the well-known result of Morales and O'Neil \cite{MoralesONeil:1972} and Dewar \cite{Dewar:1972}, 
$\delta\omega \sim \sqrt{\phi_1} f''(\bar{\act})$, and to our theory for three different values of 
$k\lambda_D \equiv \sigma$: 0.4 (a), 0.3 (b), and 0.2 (c).  For the Dewar/O'Neil theory, we use equation (\ref{eqn:SmMean}) which is of the form first derived by Morales and O'Neil \cite{MoralesONeil:1972}, but with the $O(1)$ pre-factor of Dewar \cite{Dewar:1972}.  We evaluate the action \emph{ansatz} value of 
$\delta\omega$ by numerically integrating (\ref{eqn:FreqShift}), with the average action $\bar{\act}$ given by the relations (\ref{eqn:dielectric1}) and (\ref{eqn:MeanAct}).

As we can see in Fig.~\ref{fig:frequency}, the Dewar/O'Neil theory agrees with our results for small values of $\phi_1$, but deviate at larger values of the potential.  Furthermore, our nonlinear theory captures both the qualitative and quantitative features of $\delta\omega$ seen in the particle simulations over the entire range $\phi_1$.  The discrepancy between theories becomes particularly clear for colder plasmas; for $k\lambda_D = 0.2$, (c) shows that the Dewar/O'Neil theory is only applicable when there is negligible frequency shift, while the action \emph{ansatz} yields reasonable qualitative agreement for all amplitudes.  We do note, however, that the quantitative agreement is worse than that observed for warmer plasmas; we speculate that this is due to the significant harmonic content of these highly nonlinear Langmuir waves.

The range of $\phi_1$ over which $\delta\omega$ was measured in Fig.~\ref{fig:frequency} includes all electrostatic amplitudes that were attained via resonant excitation with the drive amplitude $V=0.01$ and frequency equal to $\omega_L$.  Further driving of the plasma results in a ringing of $\phi_1$ that we interpret as resulting from the de-tuning of the nonlinear Langmuir wave from the external drive.

Finally, we conclude this section with simulation results regarding the measured change in the mean canonical action $\bar{\action}$.  To measure this, we use the extracted frequency shift $\delta\omega$ to determine the energy of each particle using the total potential (including harmonics): 
$H_j = \tfrac{1}{2}(p_j + \delta\omega)^2 - \phi(\zeta_j)$.  Solving this for the momentum, we numerically integrate the exact frequency-shifted action for each particle, and obtain the average $\bar{\action}$ via
\begin{align}
  \bar{\action} = \frac{1}{N}\sum_j \frac{1}{2\pi}\oint \! d\zeta \, \sqrt{2\left[ H_j + \phi(\zeta) \right]}
	- \delta\omega.		\label{eqn:simJbar}
\end{align}
In the usual adiabatic approximations, the particle action is conserved so that (\ref{eqn:simJbar}) is constant.  However, our theory predicts $\bar{\action}$ to increase as the potential grows.  We interpret this as resulting from the manner in which we excite the plasma: since the plasma is resonantly driven, the time-scale separation between the drive and the plasma response can be vanishingly small even for slowly changing amplitudes.  Thus, the particle action need not be an adiabatic invariant.

\begin{figure}
\begin{centering}
\includegraphics[scale=.53]{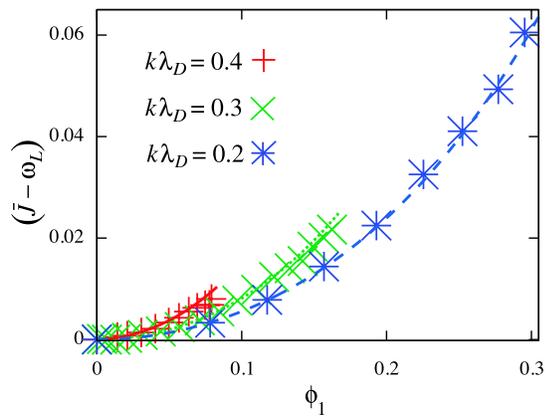}
\caption{Change in the canonical action $\bar{\action}$ for three different temperatures: $k\lambda_D=0.4$, $k\lambda_D=0.3$, and $k\lambda_D=0.2$.  The points are obtained from the simulation while the curves are numerical evaluations of $\bar{\action}$ using $\bar{\action}\equiv \bar{\act}-\delta\omega$.  As in Fig.~\ref{fig:frequency}, the range of $\phi_1$ includes all amplitudes attained with a $V=0.01$ resonant drive; we interpret the saturation of $\phi_1$ to be caused by de-tuning from the drive.}		\label{fig:simJ}
\end{centering}	
\end{figure}

To further validate this claim, we present the simulation results for the change in the average canonical action $\bar{\action}$ in Fig.~\ref{fig:simJ}, and compare this to our theory for the temperatures of Fig.~\ref{fig:frequency}.  We see remarkable agreement for temperatures such that $k_2\lambda_D = 0.4,\,0.3,\,0.2$.  First, this indicates that the canonical action of the particles is not conserved in these slowly-driven problems.  Rather, we find a well-described relation between the  electrostatic potential and the change in the average action.  In the small amplitude limit, the change in 
$\bar{\action}$ scales as $\phi_1^{3/2}$ as shown in the Appendix (\ref{eqn:Jshift}), while numerical evidence indicates that it scales roughly as $\phi_1^2$ for $\phi_1 \gtrsim \sigma^2$.

\section{Conclusion}		\label{section:conclusions}

We have presented a nonlinear electron distribution function that naturally arises for slowly driven Langmuir waves, and is parameterized by the amplitude of the electrostatic potential.  The distribution function is described by a gaussian in the canonical action whose mean varies self-consistently according to the slowly evolving potential, and whose variance remains fixed at the initial plasma temperature.  We then determined the nonlinear frequency shift of the Langmuir wave 
(\ref{eqn:FreqShift}), which agrees well with full particle simulations.  This frequency shift could be used in a reduced, fluid-like model for driven plasmas, while the asymptotic distribution function 
(\ref{eqn:f_kappa}) hints at a simplified model of nonlinear Landau damping via the dynamical process of phase-mixing.  Present research aims to use these results as the kinetic foundation of an extended three-wave model of Raman backscatter in a thermal plasma.

\acknowledgments
The authors would like to acknowledge useful discussions with R.L.~Berger  and D.J.~Strozzi.  This work was supported by the University Education Partnership Program through the Lawrence Livermore National Laboratory and DOE Grant No.~DE-FG02-04ER41289. 

\appendix*
\section{Linear and small amplitude integrals for $\bar{\act}$, $\delta\omega$}

To evaluate the linear and small amplitude limits of the dielectric $\Ez(\omega_L,\phi_1)$, we first integrate (\ref{eqn:dielectric1}) by parts.  The boundary terms at $\kappa = 0,\,\pm \infty$ vanish, while those at $\kappa=1$ cancel, resulting in the following formula for the nonlinear dielectric:
\begin{align}
\begin{split}
   \Ez (\omega_L,\phi_1) &= 1 + \int\limits_1^\infty \! d\kappa \; \frac{h(\kappa)}{\act_s}\left\{
   	f_\text{I}(\kappa)\Big[ \act(\kappa) - \omega_L \Big] \right.  \\
  &\phantom{=|\sigma^2 + \int\limits_1^\infty \! d\kappa \; h(| } 
  	\left. + f_\text{III}(\kappa)\Big[ \act(\kappa) + \omega_L \Big] \right\}  \\
  &\phantom{=|} + \int\limits_0^1 \! d\kappa \; \frac{q(\kappa)}{\act_s} \left\{
  	f_\text{II}^{-}(\kappa) \Big[ \act(\kappa) - \omega_L \Big] \right.  \\
  &\phantom{=| + \int\limits_1^\infty \! d\kappa \; h(| m}  
  	\left. + f_\text{II}^{+}(\kappa)\Big[ \act(\kappa) + \omega_L \Big] \right\},	\label{eqn:dielectric2}
\end{split}
\end{align}
where
\begin{align*}
  h(\kappa) &\equiv \frac{32}{3\pi^2\sigma^2}\Big[(2\kappa^3-\kappa)\mathscr{E}(1/\kappa) - 
  	2(\kappa^3-\kappa)\mathscr{K}(1/\kappa)\Big]  \\
  q(\kappa) &\equiv \frac{32}{3\pi^2\sigma^2}\Big[(2\kappa^2 - 1)\mathscr{E}(\kappa) - 
  	(\kappa^2-1)\mathscr{K}(\kappa)\Big].
\end{align*}
First, we calculate (\ref{eqn:dielectric2}) in the linear limit, for which $\phi_1\rightarrow 0$.  In this case, the integrals over the trapped electrons (for which $0 \le \kappa \le 1$) make no contribution, while the remaining integrand vanishes exponentially for small $\kappa$.  Thus, we consistently take $\kappa \gg 1$, for which we have
\bse	 	  	\label{eqn:LargeKap}
\begin{align}
  \frac{h(\kappa)}{\act_s} f_\text{I}(\kappa) &\approx \left[\frac{1}{\kappa} + O\!\left(1/\kappa^3\right)\right]
  	\frac{e^{-\tfrac{1}{2\sigma^2}\left[\act(\kappa)-\bar{\act}\right]^2}}{\sqrt{2\pi}\sigma^3} 
	\vphantom{\int\limits_0} \\
  	 \act(\kappa) &\equiv \act_s \kappa\mathscr{E}(1/\kappa) \approx 2\kappa\sqrt{\phi_1} + O(1).
\end{align}
\ese
Defining the variables
\begin{align*}
  \sigma x &\equiv 2\kappa\sqrt{\phi_1} + \omega_L &
  	\sigma y &\equiv -2\kappa\sqrt{\phi_1} + \omega_L,
\end{align*}
and using the large $\kappa$ relations (\ref{eqn:LargeKap}), we find that the linear relation $\displaystyle{\lim_{\phi_1\rightarrow 0} \Ez(\omega_L,\phi_1)}=0$ is given by
\be
\begin{split}
  0 &= 1 + \frac{1}{\sigma^2} + \frac{\omega_L/\sigma}{\sqrt{2\pi}\sigma^2}\lim_{\delta\rightarrow 0}\left\{ \,
  	\int\limits_{\tfrac{\omega_L}{\sigma}+\delta}^\infty 
  	\!\!\!\!\! dx\, \frac{e^{-x^2/2}}{x-\omega_L/\sigma} \,  \right. \\
  &\phantom{=|\sigma^2 + 1 + \frac{\omega_L/\sigma}{\sqrt{2\pi}} \lim_{\delta\rightarrow 0}mn} \left. + \!\!\!
	\int\limits_{-\infty}^{\tfrac{\omega_L}{\sigma}-\delta} \!\!\!\!\! dy\, 
	\frac{e^{-y^2/2}}{y-\omega_L/\sigma}\right\},	\label{eqn:limDispP}
\end{split}
\ee
where $\delta \equiv 2\sqrt{\phi_1}$.  As we can see, the assumed distribution function yields a prescription for treating the pole at $x=\omega_L/\sigma$: the symmetric limit is merely the principal value.  Note that this is the standard pole occurring when the particle velocity equals the phase velocity of the wave (i.e., when the particle action equals that of the separatrix defined by the infinitesimal wave), and the symmetric limit results in the Vlasov-type dispersion relation (\ref{eqn:DerivedDisp}).  Furthermore, differentiation yields the quantity $\tfrac{\partial}{\partial\omega_L}\Ez(\omega_L,0)$ relevant for the calculation of the change in the mean action.  Using the dispersion relation 
(\ref{eqn:DerivedDisp}), we find
\begin{align}
  \lim_{\phi_1 \rightarrow 0}\frac{\partial}{\partial\omega_L}\Ez(\omega_L,\phi_1) 
  &= \frac{1}{\omega_L\sigma^2} \left(\omega_L^2 - 1 - \sigma^2\right).	\label{eqn:SmDdielectric}
\end{align}

We continue by calculating the change in $\bar{\act}$ (\ref{eqn:MeanAct}) induced by the near-resonant particles, assuming the amplitude of the potential $\phi_1$ is small.  To make the integrals defining $\Ez(\omega_L,\phi_1)$ manifestly convergent, we start by first ``rewriting the 1'' in the expression for the dielectric (\ref{eqn:dielectric2}).  Assuming  the linear dispersion relation (\ref{eqn:DerivedDisp}) is satisfied, in terms of the energy $\kappa$ we have
\be
  1 =  - \mathscr{P}\!\!\int\limits_{-\infty}^\infty \!\! d\kappa \left[2\kappa\sqrt{\phi_1} 
  	- \omega_L\right]\frac{e^{-\frac{1}{2\sigma^2}
	\left(2\sqrt{\phi_1}\kappa-\omega_L\right)^2}}{\kappa\sqrt{2\pi}\sigma^3}.		\label{eqn:Disp1}
\ee
By appropriate choice of signs for $\kappa$, we can express (\ref{eqn:Disp1}) as a sum of integrals whose limits are such that $1 \le \kappa < \infty$ or $0 \le \kappa \le 1$, that we then use to replace the 1  in the nonlinear dielectric (\ref{eqn:dielectric2}).  Thus, the nonlinear dielectric is given by
\begin{align}
\begin{split}
  \Ez(\omega_L,\phi_1) &=\int\limits_1^\infty \! d\kappa 
  	\Big[ \act(\kappa) - \omega_L \Big] h(\kappa) f_\text{I}(\kappa) \\
  &\phantom{=}+\int\limits_1^\infty \! d\kappa 
  	\Big[ \act(\kappa) + \omega_L \Big] h(\kappa) f_\text{III}(\kappa) \\ 
  &\phantom{=}+\int\limits_0^1 \! d\kappa 
  	\Big[ \act(\kappa) -\omega_L\Big] q(\kappa) f_\text{II}^{-}(\kappa) \\
  &\phantom{=}+\int\limits_0^1 \! d\kappa 
  	\Big[ \act(\kappa) + \omega_L\Big] q(\kappa) f_\text{II}^{+}(\kappa) + 1, 
		\label{eqn:longIntegral}
\end{split}
\end{align}
where the number 1 is to be interpreted as the sum of integrals from (\ref{eqn:Disp1}).  In this manner, the expression (\ref{eqn:longIntegral}) is perfectly well-defined in the small amplitude limit, and this limit is simple to calculate numerically.  Taylor expanding the integrals with $1 \le \kappa < \infty$, we have
\beu
\begin{split}
  &\sqrt{\phi_1}\;\frac{256}{3\pi^2}\int\limits_1^\infty \! d\kappa \, \left[ \frac{3\pi^3}{64}  - 
	h(\kappa)\kappa\ellipticE(1/\kappa)\ellipticK(1/\kappa) \right] \\
  &\times \! \left(\tfrac{\omega_L^2}{\sigma^2} - 1\right) \!
  	\frac{e^{-\omega_L^2/2\sigma^2}}{\sigma\sqrt{2\pi}}
	\approx -1.50
	\sqrt{\phi_1} \left( \tfrac{\omega_L^2}{\sigma^2} - 1 \right)
  	\frac{e^{-\omega_L^2/2\sigma^2}}{\sqrt{2\pi}\sigma^3},
\end{split}
\eeu
while the last two integrals of (\ref{eqn:longIntegral}) yield
\beu
\begin{split}
  \sqrt{\phi_1}  \! \int\limits_0^1 \! d\kappa \left\{ 1 - \frac{64\kappa}{3\pi^3} q(\kappa)\ellipticK(\kappa)
  	\big[ \ellipticE(\kappa) + (\kappa^2-1)\ellipticK(\kappa)\big] \right\} \\
  \times\, 4 \left( \tfrac{\omega_L^2}{\sigma^2} - 1 \right) \! 
  	\frac{e^{-\omega_L^2/2\sigma^2}}{\sigma\sqrt{2\pi}} \approx 2.59
	\sqrt{\phi_1} \left( \tfrac{\omega_L^2}{\sigma^2} - 1 \right)
  	\frac{e^{-\omega_L^2/2\sigma^2}}{\sqrt{2\pi}\sigma^3}.
\end{split}
\eeu
Adding these contributions, we find that the small amplitude behavior of the nonlinear dielectric is
\begin{align}
  \lim_{\phi_1\rightarrow 0} \Ez(\omega_L,\phi_1) &= 1.089\sqrt{\phi_1} 
  	\left( \frac{\omega_L^2}{\sigma^2} - 1 \right)
	\frac{e^{-\omega_L^2/2\sigma^2}}{\sqrt{2\pi}\sigma^3}.
\end{align}

Finally, we calculate the difference between the frequency shifted average action and the frequency, namely, the average action $\bar{\action}$.  
Using the definition $\bar{\act}\equiv \left\langle \act \right\rangle$ and the expression for the frequency shift 
(\ref{eqn:FreqShift}), we find that
\begin{align*}
\begin{split}
  \bar{\act} - \big(\omega_L+\delta\omega\big) &= \int\limits_1^\infty \!\! d\kappa \left[\act(\kappa) - 
  	\frac{\pi\kappa\sqrt{\phi_1}}{\ellipticK(1/\kappa)} \right] \!
  	\left[f_\text{I}(\kappa) - f_\text{III}(\kappa) \right]  \\
  &\phantom{=|} + \int\limits_0^1 \!\! d\kappa \; \act(\kappa)
  	\left[f_\text{II}^{-}(\kappa) - f_\text{II}^{+}(\kappa) \right] .
\end{split}
\end{align*}
It can be shown that both integrals vanish in the limit $\phi_1\rightarrow 0$.  Taylor expanding the integrals for small values of the potential $\phi_1$, we find that
\begin{align}
\begin{split}
  &\bar{\act} - \big(\omega_L+\delta\omega\big) \approx \frac{32}{\pi}
  	\frac{\omega_L}{\sigma^2} \frac{e^{-\omega_L^2/2\sigma^2}}{\sigma\sqrt{2\pi}} \phi_1^{3/2} \\
  &\phantom{mmm} \times \! \left\{ \int\limits_1^\infty \!\! d\kappa \; \ellipticE(1/\kappa)\left[ 
  	\frac{4\kappa^2}{\pi^2}\ellipticE(1/\kappa)\ellipticK(1/\kappa) - \kappa^2 \right] \right. \nonumber \\
  &\phantom{mmm\times\{\{\!} + \left. \int\limits_0^1 \!\! d\kappa \; \frac{4\kappa}{\pi^2}\ellipticK(\kappa)
  	\left[\ellipticE(\kappa) + (\kappa^2 - 1)\ellipticK(\kappa)\right] \right\}
\end{split} \\
  & \phantom{mm|} 
  = \frac{64}{9\pi}
  	\frac{\omega_L}{\sigma^2} \frac{e^{-\omega_L^2/2\sigma^2}}{\sigma\sqrt{2\pi}} \phi_1^{3/2}.
		\label{eqn:Jshift}
\end{align}
Thus, we see that the average action $\bar{\action}$ grows as $\phi_1^{3/2}$ for $\phi_1$ small.  For Langmuir wave amplitudes such that these terms can be neglected, the frequency shift is equal to the change in the frequency shifted action, and the canonical action is approximately conserved.



\end{document}